\documentclass{article}

\usepackage{arxiv}

\usepackage[utf8]{inputenc} 
\usepackage[T1]{fontenc}    
\usepackage{hyperref}       
\usepackage{url}            
\usepackage{booktabs}       
\usepackage{amsfonts}       
\usepackage{nicefrac}       
\usepackage{microtype}      
\usepackage{lipsum}		
\usepackage{graphicx}
\usepackage{cite}
\usepackage{doi}
\usepackage{amsmath}
\usepackage{multirow}
\usepackage{tipa}

\title{Audio–Vision Contrastive Learning for Phonological Class Recognition}

\author{
\href{https://orcid.org/0009-0001-0905-3833}{\includegraphics[scale=0.06]{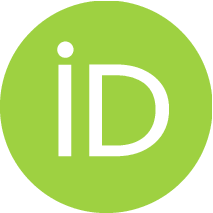}\hspace{1mm}Daiqi Liu}\thanks{Corresponding author: daiqi.deutschfau.liu@fau.de} \quad
\href{https://orcid.org/0000-0001-9405-4154}{\includegraphics[scale=0.06]{orcid.eps}\hspace{1mm}Tomás Arias-Vergara} \quad
\href{https://orcid.org/0000-0003-3476-3500}{\includegraphics[scale=0.06]{orcid.eps}\hspace{1mm}Jana Hutter} \quad
\href{https://orcid.org/0000-0000-0000-0000}{\includegraphics[scale=0.06]{orcid.eps}\hspace{1mm}Andreas Maier} \quad
\href{https://orcid.org/0000-0002-2727-2116}{\includegraphics[scale=0.06]{orcid.eps}\hspace{1mm}Paula Andrea Pérez-Toro} \\
Pattern Recognition Lab, Friedrich-Alexander-Universität Erlangen-Nürnberg, Germany \\
Smart Imaging Lab, Friedrich-Alexander-Universität Erlangen-Nürnberg, Germany \\
GITA Lab, Universidad de Antioquia UdeA, Medellín, Colombia \\
\texttt{daiqi.deutschfau.liu@fau.de} \\
\texttt{tomas.arias@fau.de} \\
\texttt{jana.hutter@fau.de} \\
\texttt{andreas.maier@fau.de} \\
\texttt{paula.andrea.perez@fau.de}
}

\date{}

\renewcommand{\headeright}{Daiqi Liu et al.}
\renewcommand{\undertitle}{}
\renewcommand{\shorttitle}{Audio–Vision Contrastive Phonology}

\hypersetup{
pdftitle={Audio–Vision Contrastive Learning for Phonological Class Recognition},
pdfauthor={Daiqi Liu, Tomás Arias-Vergara, Jana Hutter, Andreas Maier, Paula Andrea Pérez-Toro},
}

\begin{document}
\maketitle

\begin{abstract}
Accurate classification of articulatory-phonological features plays a vital role in understanding human speech production and developing robust speech technologies, particularly in clinical contexts where targeted phonemic analysis and therapy can improve disease diagnosis accuracy and personalized rehabilitation. 
In this work, we propose a multimodal deep learning framework that combines real-time magnetic resonance imaging (rtMRI) and speech signals to classify three key articulatory dimensions: manner of articulation, place of articulation, and voicing.
We perform classification on 15 phonological classes derived from the aforementioned articulatory dimensions and evaluate the system with four audio/vision configurations: unimodal rtMRI, unimodal audio signals, multimodal middle fusion, and contrastive learning-based audio-vision fusion. 
Experimental results on the USC-TIMIT dataset show that our contrastive learning-based approach achieves state-of-the-art performance, with an average F1-score of 0.81, representing an absolute increase of 0.23 over the unimodal baseline. The results confirm the effectiveness of contrastive representation learning for multimodal articulatory analysis. Our code and processed dataset will be made publicly available at \url{https://github.com/DaE-plz/AC_Contrastive_Phonology} to support future research.
\keywords{Real-Time MRI, Contrastive Learning, Multimodal Deep Learning, Phoneme Recognition}
\end{abstract}

\section{Introduction}
Understanding the articulatory processes underlying human speech production is a long-standing challenge in speech science, with profound implications for theoretical linguistics and clinical diagnostics~\cite{franciscatto2021towards}. Real time magnetic resonance imaging (rtMRI) has emerged as a powerful tool to visualize the dynamic configurations of the vocal tract during speech production, offering unparalleled insights into articulatory mechanisms~\cite{toutios2016advances}. 
Such insights can be quantified through phonemic analysis, enabling deeper understanding of human speech.
However, extracting meaningful phonemic representations from these high-dimensional and often noisy sequences remains challenging.
On the one hand, using only MRI without audio recordings limits the system's ability to accurately distinguish between certain phonemes, mainly because of the subtle changes in the surrounding vocal tract structures during sound coarticulation~\cite{aleksic2009audiovisual}; thus, integrating speech recordings during MRI scans enhances phonemic recognition accuracy. 
On the other hand, acquiring simultaneous speech recordings requires the use of non-magnetic microphones (e.g., fiber-optic) which might not be available due to the high cost of such devices.  
Therefore, methodologies enabling the integration of audio-visual data acquired during MRI scans for phonemic analysis are required.

\subsection{Related Work}

\begin{table}[ht]
\centering
\caption{Phoneme classes considered in this study.}
\label{tab:phoneme_classes}
\resizebox{\linewidth}{!}{
\begin{tabular}{llll}
\hline \hline
 \textbf{Dimension} & \textbf{Class} & \textbf{Phonemes} & \textbf{Brief description} \\
\hline
 -        & Silence      & --                                         & Non-speech segments \\
\hline
 Manner   & Stop         & \textipa{/p/, /t/, /k/, /b/, /d/, /g/}     & Total oral closure with rapid release \\
   & Nasal        & \textipa{/n/, /m/, /N/}                    & Airflow through the nasal cavity \\
    & Fricative    & \textipa{/s/, /S/, /z/,  /f/}          & Hissing sounds due to turbulent airflow \\
    & Approximants & \textipa{/j/}                              & Hissing sounds without turbulent airflow \\
    & Vowel        & \textipa{/a/, /e/, /i/, /o/, /u/}          & Vibration of the vocal folds \\
\hline

 Place    & Labial       & \textipa{/p/, /b/, /m/, /f/, /v/}          & Lips and teeth \\
  & Dental       & \textipa{/T/, /D/}          & Tongue against upper teeth \\
     & Alveolar     & \textipa{/t/, /d/, /n/}                    & Tip of the tongue and alveolar ridge \\
    & Postalveolar & \textipa{/S/}                              & Blade of the tongue \\
     & Palatal      & \textipa{/j/}                              & Front of the tongue and hard palate \\
    & Velar        & \textipa{/k/, /g/, /N/}                    & Back of the tongue and soft palate \\
       & Glottal      & \textipa{/h/}               & Airflow through the open glottis \\

\hline
 Voicing  & Voiceless    & \textipa{/p/, /t/, /k/,  /S/, /s/}     & No vibration of the vocal folds \\
   & Voiced       & \textipa{/m/, /n/, /b/, /d/, /g/, /a/} & Vibration of the vocal folds \\
\hline \hline
\end{tabular}}
\end{table}

Some works in the literature have use rtMRI with synchronized audio signals for phoneme recognition.
Narayanan et al.~\cite{narayanan2014real} proposed a multimodal real-time MRI articulatory corpus for speech research, providing synchronized speech and rtMRI data to facilitate the study of speech production. 
Saha et al.~\cite{saha2018towards} proposed using long-term recurrent convolutional Networks models, to identify different vowel-consonant-vowel (VCV) sequences from dynamic shaping of the vocal tract, where an accuracy of 42\% was reported in the prediction of 51 different VCV combinations.
Van Leeuwen et al.~\cite{van2019cnn} proposed a deep learning model to classify 27 different phonemes using midsagittal MRI of the vocal tract using a convolutional neural network (CNN) was trained to classify vowels (13 classes), consonants (14 classes), and phonemes (27 classes) across 17 subjects, yielding accuracies of up to 57\,\%.
Pandey et al.~\cite{pandey2021silent} combined 3D convolutional layers, bidirectional recurrent networks, and connectionist temporal classification loss to generate text from articulatory motions captured from MRI data, achieving a phoneme error rate of 40.6\% at sentence-level.

Recent advancements have focused on integrating multimodal data sources, combining acoustic signals with articulatory information to improve phoneme recognition. Contrastive learning approaches, such as the contrastive token-acoustic pre-training (CTAP) method, align phoneme and speech representations in a joint multimodal space, facilitating better frame-level connections between modalities~\cite{qiang2024learning}. 
Additionally, models like SCaLa leverage supervised contrastive learning to enhance phonemic representation learning for end-to-end speech recognition systems~\cite{fu2021scala}.
Arias-Vergara et al.~\cite{arias2024contrastive} proposed to use a contrastive learning approach with vision transformer (ViT) and Wav2Vec encoders to classify 9 phonological classes (at frame level) achieving an average F1-score of 0.85.
In that work, the authors used VCV combinations to perform the classification task.

These developments underscore a shift towards more sophisticated, data-driven approaches that leverage both acoustic and articulatory information, as well as advanced learning techniques, to achieve more accurate and robust phoneme recognition systems.

\subsection{Contributions}

Articulatory data provides a direct window into the physical processes underlying phonemic distinctions, enabling a deeper understanding of how phonemes are produced and differentiated. We present a comprehensive framework comparing ViT and Wav2Vec as encoders for three key articulatory classification tasks on sentences from the USC-TIMIT dataset: manner (6-class), place (8-class), and voicing (3-class)~\cite{vaswani2017attention,baevski2020Wav2Vec,tom2022Wav2Vec,narayanan2014usc}.
Our study systematically evaluates four modality configurations: unimodal speech, unimodal video, multimodal fusion, and contrastive learning—across both pre-trained and fine-tuned regimes. 
Experimental results demonstrate that our contrastive multimodal approach achieves 0.81 average F1-score, outperforming unimodal baselines by $\sim$23$\%$ on classification. By focusing on the classification of manner, place, and voicing, our work lays a foundational step toward more granular, generalizable, and interpretable phoneme recognition. 

\section{Datasets}

\begin{table}[ht]
\centering
\caption{Number of frames per phonological class after labeling}
\label{tab:phonological_class_counts}
\resizebox{0.94\linewidth}{!}{
\begin{tabular*}{0.85\textwidth}{@{\extracolsep{\fill}}lcc}
\hline \hline
\textbf{Phonological Class} & \textbf{Number of Frames} & \textbf{Percentage} \\
\hline 
Silence        & 141{,}013 & 29.04\% \\
Voiced         & 114{,}623 & 23.60\% \\
Voiceless      & 24{,}158  & 4.97\%  \\
Labial         & 9{,}963   & 2.05\%  \\
Dental         & 2{,}147   & 0.44\%  \\
Alveolar       & 40{,}934  & 8.43\%  \\
Postalveolar   & 4{,}034   & 0.83\%  \\
Palatal        & 985       & 0.20\%  \\
Velar          & 7{,}684   & 1.58\%  \\
Glottal        & 1{,}267   & 0.26\%  \\
Stop           & 17{,}531  & 3.61\%  \\
Nasal          & 11{,}000  & 2.27\%  \\
Fricative      & 21{,}972  & 4.52\%  \\
Approximant    & 16{,}511  & 3.40\%  \\
Vowel          & 71{,}767  & 14.78\% \\
\hline \hline
\end{tabular*}}
\end{table}

This study employs the USC-TIMIT database, a multimodal corpus designed to facilitate research on human speech production using rtMRI~\cite{narayanan2014usc,klumpp2022common}. The dataset consists of recordings from 10 native speakers of American English (5 male, 5 female), each producing 460 phonetically balanced sentences.
MRI and speech data were acquired using a commercial 1.5T MRI scanner. A body coil was used for radiofrequency transmission, while signal reception was performed using a custom 4-channel upper-airway receiver array, composed of two anterior and two posterior elements. The data was collected using a 13-interleaf spiral gradient echo sequence (TR = 6.164 ms, flip angle = 15°, FOV = $200 \times 200$ mm). Each mid-sagittal image slice had a thickness of 5 mm and a resolution of $68 \times 68$ pixels (2.9 mm isotropic). The final frame rate was approximately 23.18 frames per second, achieved via view-sharing reconstruction using a sliding window approach. Synchronized speech was simultaneously recorded using a fiber-optic microphone at a sampling rate of 20~kHz. We down-sampled the speech signals to meet the specifications of the speech encoder used in  this study.

\subsection{Phonological Labeling}

Table~\ref{tab:phoneme_classes} shows the three articulatory-phonological groups considered in our study, along with the mapping from phonemes to each group. 
The audio signal's phonetic transcriptions were provided with the dataset.
Phonemes were labeled according to the ARPABET.
Following standard phonetic tables, we grouped the phonemes according to \textit{manner of articulation}, \textit{place of articulation}, and \textit{voicing}.
The resulting time stamps were then used to assign frame-level labels (to the audios and MRIs) by aligning them with the temporal indices of the MRI and speech streams. For non-speech segments, we performed manual corrections and labeled the audio/MRI frames as silence. As shown in Table~\ref{tab:phonological_class_counts}, the resulting dataset exhibits a naturally imbalanced distribution across phonological classes, with silence and vowels comprising the largest proportions of labeled frames.

\section{Methods}

Our framework addresses frame-level phonological classification using synchronized rtMRI images and speech signals. During training, a single video frame and its corresponding speech signal are processed by two different encoders. The image encoder is fine-tuned, while the parameters of the speech encoder remain unchanged. Before passing the MRI feature embedding for classification,we maximize the similarity between the image and speech embeddings. For this, we must first project the MRI embedding
into the same dimensions as the audio encoder. Then, a linear multilayer perceptron with a softmax activation function is used for the classification. Besides the contrastive loss, we incorporate a class-balanced, learnable weighting scheme into the cross-entropy loss. which is designed to mitigate class imbalance by assigning higher importance to underrepresented phonological classes during training. Finally, we combine them to improve classification accuracy. During inference, only the MRI is used to predict the phonological classes.

\subsection{Processing}

We resampled the MRI videos to 15~fps. Each grayscale MRI frame was resized to a resolution of $128 \times 128$ pixels to match the input requirements of the ViT encoder~\cite{dosovitskiy2020image}, and then duplicated across three channels to conform to the expected input format of ViT models. The speech signal was resampled to 16~kHz and subsequently padded to a fixed-length window, ensuring uniform input dimensions for Facebook's Wav2Vec base model~\cite{baevski2020Wav2Vec}. 

Frame-level classification was performed by aligning each MRI frame with a corresponding segment of speech centered around its timestamp. Specifically, a fixed-size temporal window was applied around each frame's time point to extract the associated speech segment, enabling synchronized multimodal input for training. As a result, each image frame corresponds to approximately 66.67\,ms of speech.

\subsection{Image Encoder}

MRI frames are processed using a ViT encoder implemented via the MONAI library~\cite{dosovitskiy2020image}\footnote{\url{https://docs.monai.io/en/1.3.0/_modules/monai/networks/nets/vit.html}}. The model is initialized with random weights and trained end-to-end for the phonological classification task. Each input frame is divided into non-overlapping patches of size $16 \times 16$, resulting in a sequence of 196 tokens (plus one \texttt{[CLS]} token). Each patch is linearly embedded into a 768-dimensional vector, and positional encodings are added to preserve spatial information. These embeddings are then passed through a Transformer encoder comprising 12 layers, each consisting of a multi-head self-attention mechanism with 12 attention heads and a feed-forward network of dimension 3072. Layer normalization and dropout (rate 0.1) are applied after each sub-layer. In the unimodal and multimodal classification settings, only the \texttt{[CLS]} token is used as the image representation. In contrast, the contrastive learning setup retains all patch-level outputs from the final Transformer layer.

\begin{figure}
  \centering
  \includegraphics[width=\linewidth]{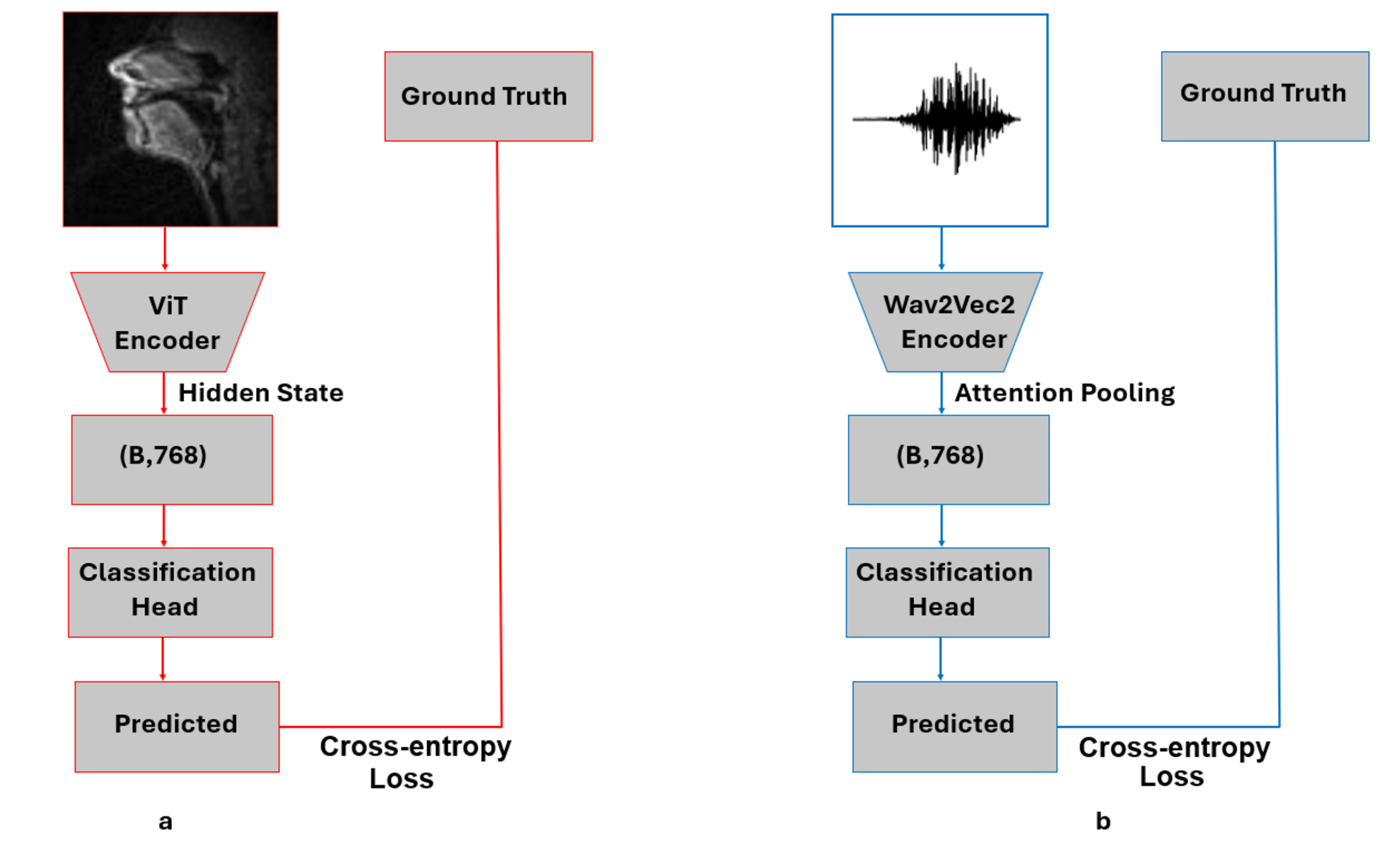}
  \caption{Overview of our proposed network for different phonological class recognizer, a) unimodal with image input (on the left), b) unimodal with audio input (on the right)}
  \label{framework1}
\end{figure}

\subsection{Speech Encoder}

\begin{figure}
  \centering
  \includegraphics[width=\linewidth]{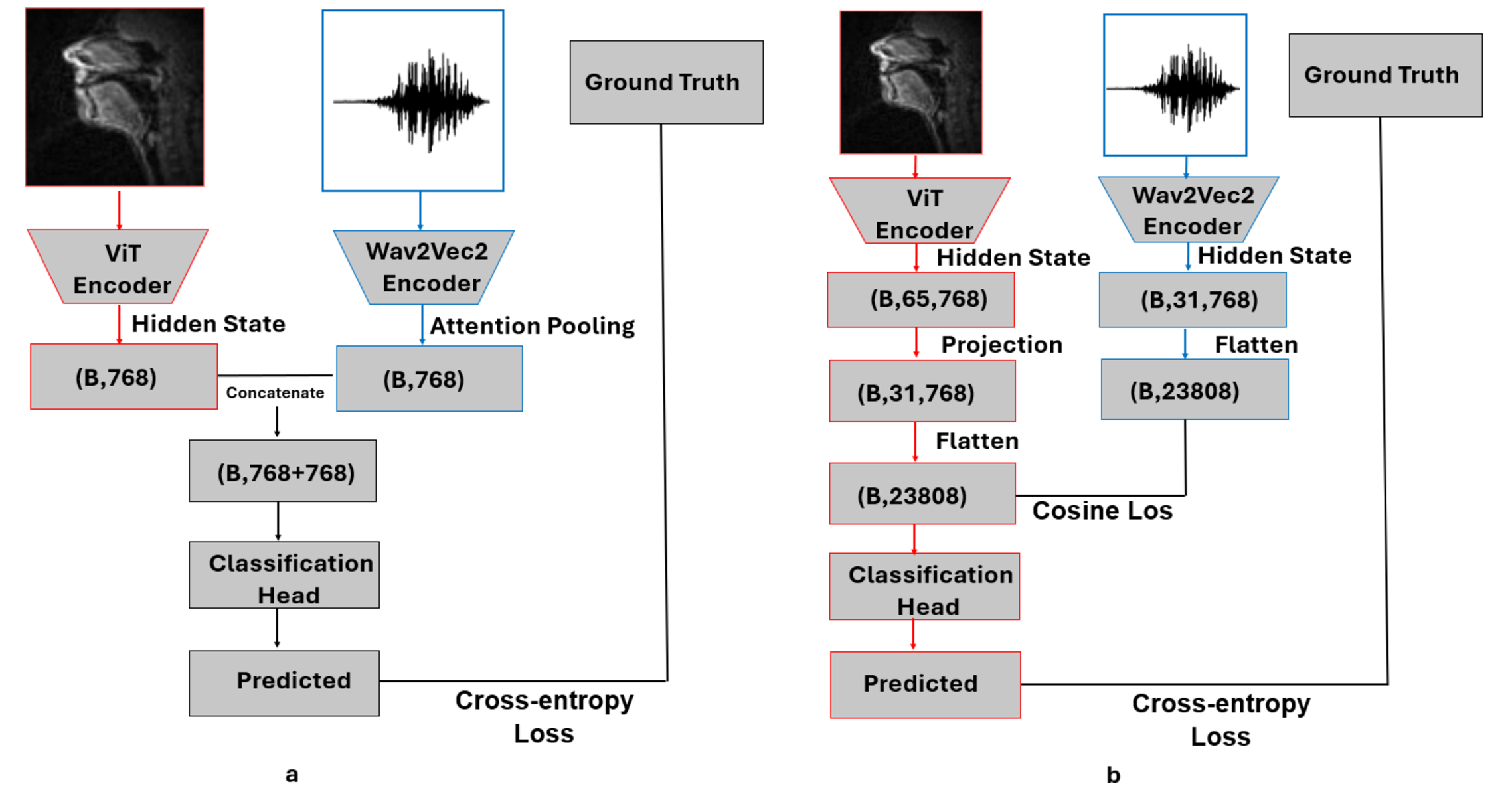}
  \caption{Overview of our proposed network for different phonological class recognizer, a) multimodal with image and audio input (on the left), b) contrastive learning model (on the right)}
  \label{framework2}
\end{figure}

We use the Wav2Vec2 base model~\cite{baevski2020Wav2Vec}\footnote{\url{https://huggingface.co/facebook/Wav2Vec2-base-960h}}, a 12-layer Transformer encoder pre-trained on 960 hours of 16~kHz speech from the LibriSpeech corpus using self-supervised learning. This base version of Wav2Vec~2.0 consists of three main components: a convolutional feature extractor, a 12-layer Transformer-based context network with 8 attention heads and a hidden size of 768, and a final projection layer. The convolutional front-end converts raw speech into a latent sequence of embeddings $z_1, \dots, z_T$, which serve as frame-level acoustic representations for downstream articulatory classification tasks in our framework. These features are used for contrastive alignment with visual features from the ViT encoder. The model uses masked and quantized segments during pre-training and produces contextualized embeddings through its Transformer layers.

\subsection{Model}

\begin{table}[ht]
\centering
\caption{Performance comparison for \textbf{manner of articulation} classification. \textbf{uniV}: unimodal model with video input. \textbf{uniA}: unimodal model with audio input.}
\label{tab:manner_results}
\resizebox{\linewidth}{!}{
\begin{tabular}{|l@{\hskip 6pt}|
                c@{\hskip 6pt}c@{\hskip 6pt}c@{\hskip 6pt}|
                c@{\hskip 6pt}c@{\hskip 6pt}c@{\hskip 6pt}|
                c@{\hskip 6pt}c@{\hskip 6pt}c@{\hskip 6pt}|
                c@{\hskip 6pt}c@{\hskip 6pt}c@{\hskip 6pt}|}
\hline
\textbf{Class} & \multicolumn{3}{c|}{\textbf{uniV}} & \multicolumn{3}{c|}{\textbf{uniA}} & \multicolumn{3}{c|}{\textbf{Multimodal}} & \multicolumn{3}{c|}{\textbf{Contrast}} \\
\hline
 & \textbf{Prec} & \textbf{Rec} & \textbf{F1} & \textbf{Prec} & \textbf{Rec} & \textbf{F1} & \textbf{Prec} & \textbf{Rec} & \textbf{F1} & \textbf{Prec} & \textbf{Rec} & \textbf{F1} \\
\hline
Silence      & 0.69 & 0.62 & 0.65 & 0.81 & 0.69 & 0.75 & 0.87 & 0.81 & 0.84 & 0.91 & 0.87 & 0.89 \\
Stop         & 0.50 & 0.55 & 0.52 & 0.60 & 0.58 & 0.59 & 0.75 & 0.72 & 0.73 & 0.85 & 0.88 & 0.86 \\
Nasal        & 0.52 & 0.50 & 0.51 & 0.60 & 0.62 & 0.61 & 0.72 & 0.70 & 0.71 & 0.85 & 0.80 & 0.82 \\
Fricative    & 0.48 & 0.50 & 0.49 & 0.55 & 0.57 & 0.56 & 0.70 & 0.68 & 0.69 & 0.80 & 0.85 & 0.82 \\
Approximant  & 0.30 & 0.35 & 0.32 & 0.40 & 0.45 & 0.42 & 0.55 & 0.50 & 0.52 & 0.65 & 0.70 & 0.67 \\
Vowel        & 0.50 & 0.54 & 0.52 & 0.55 & 0.53 & 0.54 & 0.73 & 0.69 & 0.71 & 0.85 & 0.79 & 0.82 \\
\hline
AVG & 0.50 & 0.51 & 0.52 & 0.59 & 0.57 & 0.58 & 0.72 & 0.68 & 0.71 & \textbf{0.82} & \textbf{0.81} & \textbf{0.81} \\
\hline
\end{tabular}}
\end{table}

In the unimodal image setting, each MRI frame is processed using a ViT encoder, and the \texttt{[CLS]} token from the final layer is used as the image representation (see Figure~\ref{framework1}a). In the unimodal audio setting, speech segments are encoded using Wav2Vec 2.0, and frame-level acoustic embeddings are aggregated via attention-based pooling (as illustrated in Figure~\ref{framework1}b). For the multimodal model, the \texttt{[CLS]} token from the ViT and the pooled audio embedding are concatenated and passed through a linear classification head (see Figure~\ref{framework2}a).

As illustrated in Figure~\ref{framework2}b, To enhance MRI-based classification by leveraging information from synchronized speech, we adopt a contrastive training strategy. During training, each paired MRI frame and speech segment is processed by separate encoders. The MRI input is passed through the ViT encoder, resulting in a sequence of patch-level features. These are projected through a learned linear layer to match the temporal dimension of the speech encoder output (31 speech frames). The resulting image embedding tensor is then passed through an MLP to produce projected features of shape $[\text{batch}, T, D]$, where $T$ is the number of time steps and $D$ is the projection dimension (768 in our case). The speech encoder output (also of shape $[\text{batch}, T, D]$) is similarly projected using a separate MLP. Both encoders are aligned at the temporal level.

To compute the contrastive loss, we flatten the temporal sequence of both modalities into vectors of shape $[\text{batch}, T \cdot D]$, and apply a cosine embedding loss that encourages each MRI–speech pair to have high similarity, while dissimilar pairs are implicitly pushed apart. This contrastive objective acts as an auxiliary signal alongside standard classification.

The final training loss is a weighted sum of the cross-entropy loss and the contrastive loss:
\[
\mathcal{L} = \mathcal{L}_{\text{cls}} + \lambda \cdot \mathcal{L}_{\text{cos}}
\]
where $\lambda$ is a hyperparameter (set to 0.1) controlling the influence of contrastive supervision.

\begin{table}[ht]
\centering
\caption{Performance comparison for \textbf{place of articulation} classification.}
\label{tab:place_results}
\resizebox{\linewidth}{!}{
\begin{tabular}{|l@{\hskip 6pt}|
                c@{\hskip 6pt}c@{\hskip 6pt}c@{\hskip 6pt}|
                c@{\hskip 6pt}c@{\hskip 6pt}c@{\hskip 6pt}|
                c@{\hskip 6pt}c@{\hskip 6pt}c@{\hskip 6pt}|
                c@{\hskip 6pt}c@{\hskip 6pt}c@{\hskip 6pt}|}
\hline
\textbf{Class} & \multicolumn{3}{c|}{\textbf{uniV}} & \multicolumn{3}{c|}{\textbf{uniA}} & \multicolumn{3}{c|}{\textbf{Multimodal}} & \multicolumn{3}{c|}{\textbf{Contrast}} \\
\hline
 & \textbf{Prec} & \textbf{Rec} & \textbf{F1} & \textbf{Prec} & \textbf{Rec} & \textbf{F1} & \textbf{Prec} & \textbf{Rec} & \textbf{F1} & \textbf{Prec} & \textbf{Rec} & \textbf{F1} \\
\hline
Silence      & 0.67 & 0.74 & 0.70 & 0.71 & 0.73 & 0.72 & 0.90 & 0.85 & 0.87 & 0.95 & 0.90 & 0.92 \\
Labial       & 0.77 & 0.73 & 0.75 & 0.62 & 0.68 & 0.65 & 0.90 & 0.88 & 0.89 & 0.92 & 0.90 & 0.91 \\
Dental       & 0.50 & 0.55 & 0.52 & 0.60 & 0.58 & 0.59 & 0.70 & 0.68 & 0.69 & 0.80 & 0.85 & 0.82 \\
Alveolar     & 0.55 & 0.53 & 0.54 & 0.70 & 0.68 & 0.69 & 0.80 & 0.72 & 0.76 & 0.85 & 0.70 & 0.77 \\
Postalveolar & 0.58 & 0.62 & 0.60 & 0.35 & 0.32 & 0.34 & 0.87 & 0.85 & 0.86 & 0.90 & 0.81 & 0.85 \\
Palatal      & 0.30 & 0.25 & 0.27 & 0.35 & 0.40 & 0.37 & 0.50 & 0.45 & 0.47 & 0.60 & 0.58 & 0.59 \\
Velar        & 0.40 & 0.45 & 0.42 & 0.55 & 0.53 & 0.54 & 0.65 & 0.60 & 0.62 & 0.75 & 0.80 & 0.77 \\
Glottal      & 0.25 & 0.20 & 0.22 & 0.35 & 0.30 & 0.32 & 0.50 & 0.45 & 0.47 & 0.60 & 0.55 & 0.57 \\
\hline
AVG & 0.50 & 0.54 & 0.52 & 0.56 & 0.53 & 0.54 & 0.73 & 0.69 & 0.71 & \textbf{0.80} & \textbf{0.76} & \textbf{0.78} \\
\hline
\end{tabular}}
\end{table}

\begin{table}[ht]
\centering
\caption{Performance comparison for \textbf{voicing} classification.}
\label{tab:voicing_results}
\resizebox{\linewidth}{!}{
\begin{tabular}{|l@{\hskip 6pt}|
                c@{\hskip 6pt}c@{\hskip 6pt}c@{\hskip 6pt}|
                c@{\hskip 6pt}c@{\hskip 6pt}c@{\hskip 6pt}|
                c@{\hskip 6pt}c@{\hskip 6pt}c@{\hskip 6pt}|
                c@{\hskip 6pt}c@{\hskip 6pt}c@{\hskip 6pt}|}
\hline
\textbf{Class} & \multicolumn{3}{c|}{\textbf{uniV}} & \multicolumn{3}{c|}{\textbf{uniA}} & \multicolumn{3}{c|}{\textbf{Multimodal}} & \multicolumn{3}{c|}{\textbf{Contrast}} \\
\hline
 & \textbf{Prec} & \textbf{Rec} & \textbf{F1} & \textbf{Prec} & \textbf{Rec} & \textbf{F1} & \textbf{Prec} & \textbf{Rec} & \textbf{F1} & \textbf{Prec} & \textbf{Rec} & \textbf{F1} \\
\hline
Silence    & 0.74 & 0.82 & 0.78 & 0.85 & 0.83 & 0.84 & 0.87 & 0.82 & 0.84 & 0.95 & 0.91 & 0.93 \\
Voiceless  & 0.55 & 0.50 & 0.52 & 0.65 & 0.68 & 0.66 & 0.75 & 0.72 & 0.74 & 0.85 & 0.80 & 0.82 \\
Voiced     & 0.60 & 0.65 & 0.62 & 0.70 & 0.72 & 0.71 & 0.80 & 0.78 & 0.79 & 0.90 & 0.88 & 0.89 \\
\hline
AVG & 0.63 & 0.66 & 0.64 & 0.73 & 0.74 & 0.73 & 0.81 & 0.77 & 0.79 & \textbf{0.90} & \textbf{0.86} & \textbf{0.88} \\
\hline
\end{tabular}}
\end{table}

\section{Experiments \& Results}

\begin{figure}[ht]
    \centering
    \includegraphics[width=\linewidth]{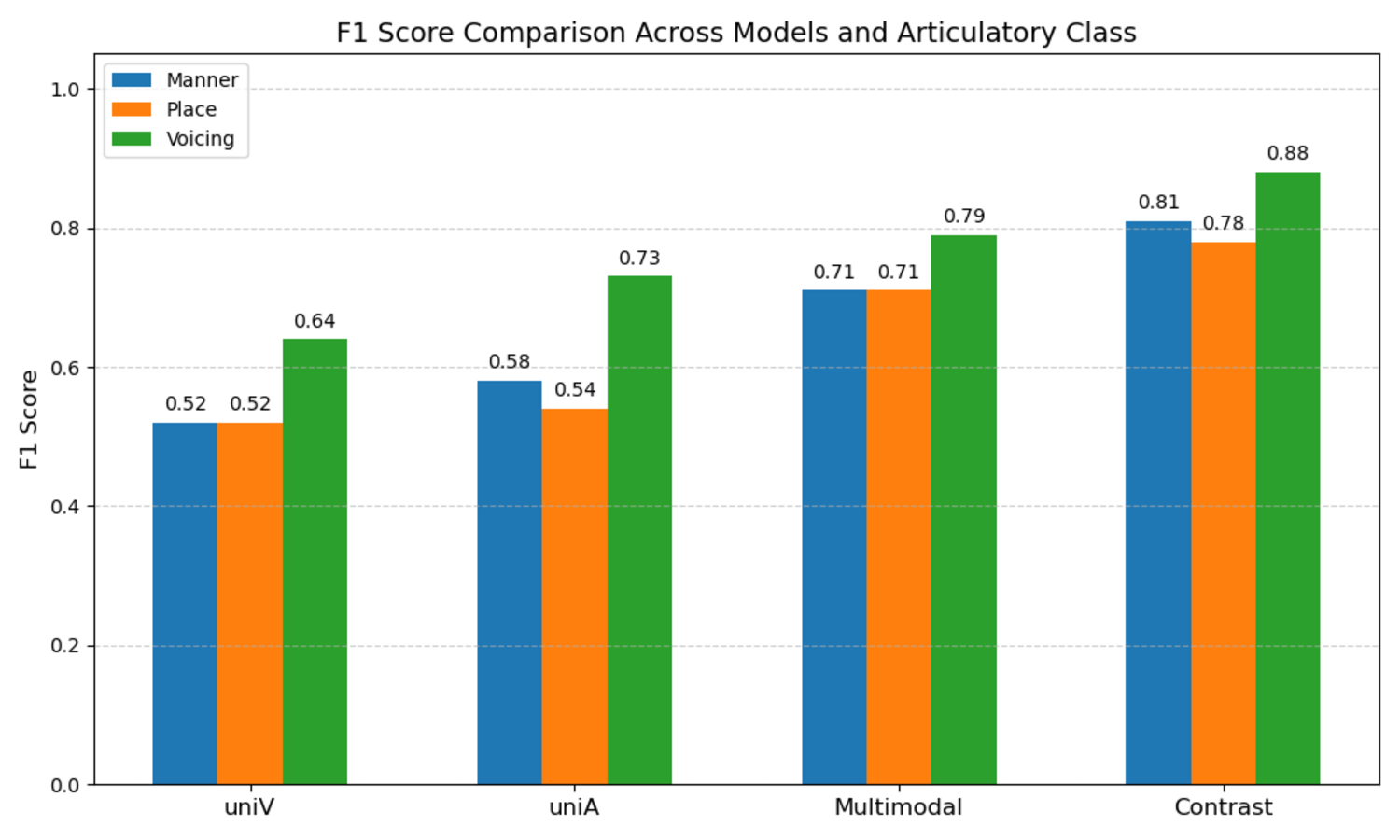}  
    \caption{Average F1-score comparison across three phonological classification tasks (manner, place, voicing) for different modality configurations.}
    \label{fig:f1_barplot}
\end{figure}

We evaluated our models using 5-fold cross-validation on the USC-TIMIT dataset to ensure robustness and generalization. In each fold, speakers were randomly divided into training (8 speakers), validation and test (2 speaker) subsets, ensuring speaker-independent splits. To mitigate gender bias, each fold maintained a balanced distribution of male and female speakers across all subsets. Model performance was assessed using frame-level Precision, Recall, and macro-averaged F1-score.

All models were trained for 30 epochs on an NVIDIA RTX A100 GPU with 40GB of memory, using a batch size of 32. The initial learning rate was set to $10^{-4}$, and weight decay to $5 \times 10^{-4}$. We used the AdamW optimizer for training stability and improved generalization.

We conducted experiments across three phonological classification tasks: (1) \textit{manner of articulation}, (2) \textit{place of articulation}, and (3) \textit{voicing}. These tasks were evaluated under four encoder-modality configurations: \textbf{uniV}, which uses only video/image data; \textbf{uniA}, which uses only audio/speech data; \textbf{Multimodal}, which combines speech and video inputs using a middle-fusion strategy; and \textbf{Contrast}, a contrastive learning approach in which speech provides complementary information to the image modality during training.

The classification was conducted in a frame-wise manner, with each MRI frame aligned to a temporally centered speech segment. The results on the test set, grouped by articulatory class, are summarized in three separate tables: Table~\ref{tab:manner_results} for manner of articulation, Table~\ref{tab:place_results} for place of articulation, and Table~\ref{tab:voicing_results} for voicing.
To provide a clearer comparison across the three phonological classification tasks, we visualized the average F1-scores of all four modality configurations (uniV, uniA, Multimodal, and Contrast) in a bar plot (see Figure~\ref{fig:f1_barplot}).

The uniV performed poorly in voicing categories such as \textit{voiced} (0.62). The \textbf{uniA} showed better performance in \textit{voiced} (0.7), \textit{voiceless} (0.66). but underperformed in place-related categories such as \textit{labial} (0.65) and \textit{alveolar} (0.69). The multimodal fusion model consistently improved F1-scores across all categories, reaching 0.89 in \textit{labial}, 0.79 in \textit{voiced}, and 0.62 in \textit{velar}. The contrastive ViT model achieved the best overall performance, with F1-scores of 0.91 in \textit{labial}, 0.89 in \textit{voiced}, and 0.77 in \textit{velar}. On the articulatory classification task, it achieved an average F1-score of 0.81, which corresponds to a 0.23 absolute F1 increase over the unimodal speech baseline, and a 0.09 improvement over the multimodal model.

\section{Discussion and Conclusions}

In this study, we propose a multimodal framework for phonological classification that integrates rtMRI and synchronized speech, aiming to systematically investigate how different modality configurations affect classification performance.

Our experiments demonstrated that while unimodal models capture modality-specific cues, they exhibit limitations when used in isolation. For instance, the unimodal video model performed suboptimally in tasks requiring fine-grained acoustic information, such as voicing classification. This limitation may be attributed to the fact that the vocal folds are not fully visible in the imaging data, making it difficult to extract the necessary articulatory cues for the voicing task~\cite{jiang2000vocal}. Conversely, the unimodal speech model was less effective for the place of articulation, particularly for visually salient articulatory patterns like labial or alveolar classes. These results underscore the complementary nature of acoustic and visual information.

By integrating both modalities, the multimodal fusion model consistently outperformed the unimodal baselines, achieving an average F1-score of 0.72 on the classification task. Furthermore, the contrastive learning framework yielded additional improvements across all three tasks, with an average F1-score of 0.81 ---representing a 23\% improvement over the best unimodal result and a 9\% gain over the multimodal fusion model. These results suggest that contrastive learning effectively enhances modality alignment for the image modality, leading to improved robustness and generalization in articulatory classification.

\begin{figure}[!htpb]
  \centering
  \includegraphics[width=0.85\linewidth]{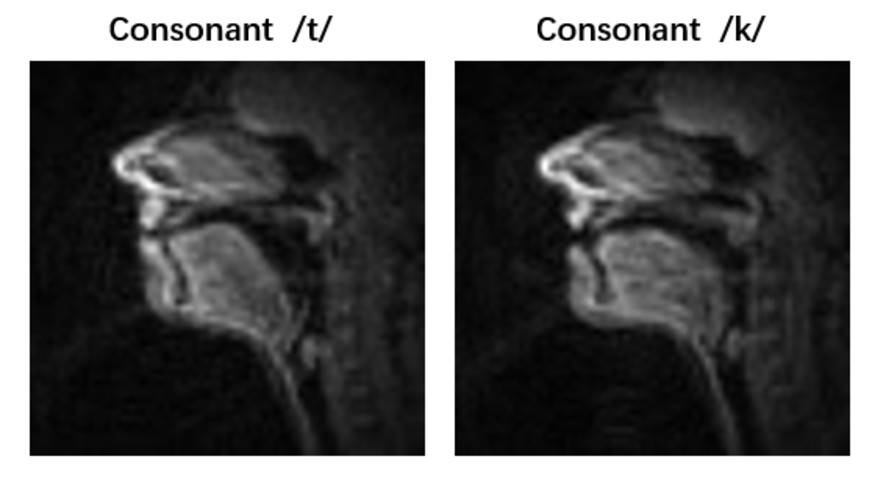}
  \caption{rtMRI frames showing tongue articulation during the production of /k/ and /t/.}
  \label{frameimg}
\end{figure}

For place of articulation, the contrastive ViT model performs worse on velar and alveolar categories compared to others, with F1-scores as low as 0.77 for both classes. The results also suggest that recognition errors may stem from articulatory similarities between phonemes with distinct places of articulation. For instance, consider Sentence~40 from the dataset: \textit{``Catastrophic economic cutbacks neglect the poor.''} This sentence contains both the /k/ and /t/ phonemes, which correspond to the \textit{velar} and \textit{alveolar} places of articulation, respectively. While these phonemes are classified differently, their articulatory gestures can be relatively close in terms of tongue placement. For instance, examining the MRI frames such as the one in Figure~\ref{frameimg}, we observed that the tongue positions for producing both consonants are similar; however, the place of articulation differs: /t/ involves the tongue tip, whereas /k/ engages the tongue body. In rtMRI, these differences may be subtle and difficult to distinguish due to limited spatial and temporal resolution. 

In addition to velar and alveolar confusions, the model performs notably poorly on glottal and palatal categories, with both classes exhibiting the lowest F1-scores among all place-of-articulation labels. This performance discrepancy is likely attributable to the severe data imbalance present in the training set. As shown in Table~\ref{tab:phonological_class_counts}, the number of annotated glottal and palatal frames constitutes only 0.26\% and 0.20\% of the total dataset, respectively. Beyond data scarcity, the inherent nature of these articulations presents additional limitations. Both palatal and glottal consonants involve articulatory gestures that are either subtle or occur in regions that are less visible in midsagittal rtMRI. For example, glottal sounds such as /h/ (as in “hat”) are produced near the vocal folds and primarily involve changes in airflow rather than large-scale tongue or jaw movements.

Despite these promising results, our current approach has several limitations. First, the model operates directly on full-frame MRI images without explicit localization or segmentation of articulators. This may limit its ability to focus on task-relevant regions, especially in noisy or anatomically variable conditions. Future work could benefit from incorporating region-of-interest (ROI) detection modules or attention mechanisms to improve interpretability and performance.

Additionally, we used pre-trained encoders with minimal domain-specific adaptation. Exploring task-specific fine-tuning strategies or domain-adaptive pre-training, particularly for visual encoders trained on MRI data, could further enhance classification accuracy. Another direction involves evaluating the framework on clinical MRI datasets, where speaker variability, pathology, and imaging conditions pose additional challenges.


\bibliographystyle{unsrt}

\end{document}